# Micelles Hydrodynamics

David V. Svintradze*

*Department of Physics, Tbilisi State University, Chavchavadze Ave. 03, Tbilisi 0128, Georgia

ABSTRACT

A micelle consists of monolayer of lipid molecules containing hydrophilic head and hydrophobic tail. These amphiphilic molecules in aqueous environment aggregate spontaneously into monomolecular layer held together due to hydrophobic effect by weak non-covalent forces. Micelles are flexible surfaces that show variety of shapes of different topology, but remarkably in mechanical equilibrium conditions they are spherical in shape. The shape and size of a micelle are functions of many variables such as lipid concentration, temperature, ionic strength, etc. Addressing the question-"why the shape of micelles is sphere in mechanical equilibrium conditions" analytically proved to be a difficult problem. In the following paper we offer the shortest and elegant analytical proof of micelles spheroidal nature when they are thermodynamically equilibrated with solvent. The formalism presented in this paper can be readily extended to any homogenous surfaces, such are vesicles and membranes.

# 1 INTRODUCTION

A micelle is an aggregate of lipid molecules dispersed in a liquid colloid. A typical micelle in aqueous solution forms an aggregate so that the hydrophilic head regions are in contact with the surrounding solvent, while the hydrophobic tail regions are pointed toward the aggregate center. The aggregation is caused by the hydrophobic and hydrophilic interactions of lipids with the surrounding water molecules [1]. In mechanical equilibrium conditions micelles are spherical in shape. The shape and size of a micelle are a function of the molecular geometry of its surfactant molecules and solution conditions, such as surfactant concentration, temperature, pH, and ionic strength. In order to answer the question-"why the shape of micelles is sphere in mechanical equilibrium conditions", it is necessary to consider the motion of micelles in fluid induced by hydrophobic and hydrophilic interactions. Therefore, the problem is to find an exact equation for micelle surface considered as fluid lipid membrane and to solve it analytically for equilibrium case.

Among the remarkable aspects of fluid lipid membranes deduced from the large body of theoretical works (see reviews [2-4]), is that the physical behavior of a membrane on the length scale not much bigger than its own thickness, can be described with high accuracy by a purely geometric Hamiltonian [5-7]. Unfortunately, this insight about curvature elastic models of membrane surfaces come along with a challenging math and have not been analytically solved. Associated Euler-Lagrange equations [8, 9], so called "shape equations", turned to be fourth order partial nonlinear differential equations, and finding a general analytical solution is a difficult problem, even though it has been numerically solved for some specific [10-17] and general cases [18, 19].

In this paper we employ different approach to the problem, namely, we use tensor calculus of moving surfaces and first law of thermodynamics to deduce the final equation for micelle dynamics and to solve it analytically for the equilibrium case.

In fluid dynamics material particles can be treated as a vertex of geometric figure and virtual layers as surfaces, and equations of motion for such surfaces can be searched. The surfaces shall be called differentially variational surfaces (DVS). We propose equations of motion of moving surfaces in aqueous solutions and apply it to analyze micelles morphology in fluid dynamics [20].

Hydrophobic and hydrophilic interactions incorporates dispersive interactions throughout the molecules, mainly related to electrostatics and electrodynamics (Van der Waals forces), induced by permanent (water molecules) or induced dipoles (dipole-dipole interactions) and possible quadrupole-quadrupole interactions (for instance stacking or London forces) plus ionic interactions (Coulomb forces). The hydrophobic effect can be considered as synonymous of dispersive interactivity with water molecules and the hydrophilic one as synonymous of polar interactivity with water molecules [21, 22]. All these interactions have one common feature and can be unified as electro-magnetic interaction's dependence on interacting bodies' geometries, where by geometries we mean shape of the objects' surfaces [20, 23]. Analytical solution of simplified DVS equations displayed all possible shapes of micelles spanning spheroids, lamellas, and cylinders. The equations can be applied to the problems related to cell motility and growth factors and show that in the mechanical equilibrium conditions with the solvent, for homogenous surfaces, a trace of the mixed curvature tensor is a pressure across the surface divided by membrane tension [20, 24]. Fully non-restrained, relativistic and exact equations for moving surfaces in electromagnetic field, when the interaction with an ambient environment is ignored, reads

$$\dot{\nabla}\rho + \nabla_i\left(\rho V^i\right) = \rho C B_i^i$$

$$\partial_\alpha\left(\rho V^\alpha\left(\dot{\nabla}C + 2V^i\nabla_i C + B_{ij}V^i V^j\right) - (\frac{1}{\mu_0}F_{\mu\nu}F^{\mu\nu} + A_\beta J^\beta)V^\alpha\right) = fa \quad (1)$$

$$\int_S \rho V_i\left(\dot{\nabla}V^i + V^j\nabla_j V^i - C\nabla^i C - CV^j B_j^i\right)dS = \int_\Omega f^i a_i d\Omega$$

where $\rho$ is the surface mass density, $V^\alpha, V^i$ are coordinate and tangential components of surface velocity, $C$ is interface velocity, $\alpha = 0,1,2,3$ for Minkowski four-dimensional space-time ambient space, $i = 0,1,2$ for pseudo-Riemannian manifold (surface), $B_{ij}$ is the surface curvature tensor, $F^{\mu\nu}$ is electromagnetic tensor, $F^\alpha = J^\alpha - \partial_\beta F^{\beta\alpha}$, $J^\alpha$ is $\alpha$ component of $\vec{J} = \{J^\alpha\}$ four current, $f, f^i$ are normal and tangential components of $\vec{F} = \{F^\alpha\}$, $a, a_i$ are the normal and tangential components of the partial time derivative of the four vector potential $\vec{A} = \{A_\alpha\}$, $S, \Omega$ stand for surface and space integrals respectively. We don't reproduce derivation of this set in this paper, rather just mention that first one is the consequence of mass conservation, second and third equations come from minimum action principle of a Lagrangian [20] and imply motion in normal direction (second equation from the set) and in tangent direction (third equation). In case of non-relativistic motion Minkowskian space becomes Euclidian, so that $\alpha = 1,2,3$ and the surface is two-dimensional Riemannian manifold $i = 1,2$. In non-relativistic frame work, after modeling the potential energy as a negative volume integral of the internal pressure and inclusion interaction with an environment, (1) further simplifies as

$$\dot{\nabla}\rho + \nabla_i\left(\rho V^i\right) = \rho C B_i^i$$

$$\partial_\alpha\left(\rho V^\alpha\left(\dot{\nabla}C + 2V^i\nabla_i C + B_{ij}V^i V^j\right) + (P^+ + \Pi)V^\alpha\right) = -\partial_\alpha(P^+ + \Pi)\cdot V^\alpha \quad (2)$$

$$\rho V_i\left(\dot{\nabla}V^i + V^j\nabla_j V^i - C\nabla^i C - CV^j B_j^i\right) = 0$$

where $P^+, \Pi$ are internal hydrodynamic and osmotic pressures, respectively (derivation of (2) is provided in the appendix). It is noteworthy that from the set (2) only second equation differs from the dynamic fluid film equations [25, 26]

$$\rho(\dot{\nabla}C + 2V^i\nabla_i C + B_{ij}V^i V^j) = \sigma B_i^i \quad (3)$$

$\sigma$ is surface tension. (3) is only valid when the surface can be described with Laplace model of surface tension [25, 26] meaning that the surface is homogeneous and the surface tension is constant, while (2) does not have that restriction. Using (3) in (2) and taking into account that in equilibrium processes internal pressure is the same as external pressure, one gets exactly the same equation of motion in normal direction as we have in this paper.

$$\partial_\alpha(\sigma V^\alpha B_i^i + (P^+ + \Pi)V^\alpha) = -\partial_\alpha(P^+ + \Pi)V^\alpha \quad (4)$$

Ideally, to address the question analytically: why the shape of micelle is sphere in thermodynamic equilibrium condition with aqueous solution, it is necessary to be derived full governing equations of motions of surfaces in solutions by including potential energy of hydrophobic-hydrophilic interactions. As it is stated above, we have already reported such equations (1) [20] and the solution of the simplified DVS equations (2) produce exactly same outcome as we have in this paper, but (1, 2) are analytically much complex to digest even though include all the information about analytical face of potential energy of hydrophobic-hydrophilic interactions and reveal hidden geometries in potential energy, as far as right hand side of second and third equations in (1) is proportional to gradient of potential energy, while left hand side caries full information about the geometric motion because it includes curvature tensor. It should be stated that (1) is fully non-restrained. It is the exact equations of motion of surfaces in electromagnetic field and upon addition an environmental interactions explain not only membrane dynamics, but also dynamics of macromolecular surfaces.

In present paper, we provide much efficient and shorter alternative way of deduction of final equation for micelles' normal motion (4) and prove it in several lines without invoking analytical face of hydration forces (1). Instead we sacrifice geometric picture of the surface full motion and geometric description of hydrophobic-hydrophilic interactions.

## 2 THEORY

### 2. 1 Basics of Differential Geometry

In this section we provide basics of tensor calculus for moving surfaces and summarize the theorems we used directly or indirectly to deduce equations for micelle dynamics. Differential geometry preliminaries we used here can be found in tensor calculus text book [26].

Suppose that $S^i$ ($i = 1, 2$) are the surface coordinates of the moving manifold (or the surface) $S$ and the ambient Euclidean space is referred to coordinates $X^\alpha$ (Figure 1). Coordinates $S^i, X^\alpha$ are arbitrarily chosen so that sufficient differentiability is achieved in both, space and time. Surface equation in ambient coordinates can be written as $X^\alpha = X^\alpha(t, S^i)$. Let the $\vec{R}$ position vector be expressed in coordinates as

$$\vec{R} = \vec{R}(X^\alpha) = \vec{R}(t, S^i) \tag{5}$$

Latin letters in indices indicate surface related tensors. Greek letters in indices show tensors related to Euclidean ambient space. All equations are fully tensorial and follow the Einstein summation convention. Covariant bases for the ambient space are introduced as $\vec{X}_\alpha = \partial_\alpha \vec{R}$, where $\partial_\alpha = \partial / \partial X^\alpha$. The covariant metric tensor is the dot product of covariant bases

$$X_{\alpha\beta} = \vec{X}_\alpha \cdot \vec{X}_\beta \tag{6}$$

The contravariant metric tensor is defined as the matrix inverse of the covariant metric tensor, so that $X^{\alpha\beta} X_{\beta\gamma} = \delta^\alpha_\gamma$, where $\delta^\alpha_\gamma$ is the Kronecker delta. As far as the ambient space is set to be Euclidian, the covariant bases are linearly independent, so that the square root of the metric tensor determinant is unit. Furthermore, the Christoffel symbols given by $\Gamma^\alpha_{\beta\gamma} = \vec{X}^\alpha \cdot \partial_\beta \vec{X}_\gamma$ vanish and set the equality between partial and curvilinear derivatives $\partial_\alpha = \nabla_\alpha$.

Now let's discuss tensors on the embedded surface with arbitrary coordinates $S^i$. Latin indexes throughout the text are used exclusively for curved surfaces and curvilinear derivative $\nabla_i$ is no longer the same as the partial derivative $\partial_i = \partial/\partial S^i$. Similar to the bases of ambient space, covariant bases of an embedded manifold are defined as $\vec{S}_i = \partial_i \vec{R}$ and the covariant surface metric tensor is the dot product of the covariant surface bases:

$$S_{ij} = \vec{S}_i \cdot \vec{S}_j \tag{7}$$

The contravariant metric tensor is the matrix inverse of the covariant one. The matrix inverse nature of covariant-contravariant metrics gives possibilities to raise and lower indices of tensors defined on the manifold. The surface Christoffel symbols are given by $\Gamma^i_{jk} = \vec{S}^i \partial_j \vec{S}_k$ and along with Christoffel symbols of the ambient space provide all the necessary tools for covariant derivatives to be defined as tensors with mixed space/surface indexes:

$$\nabla_i T^{\alpha j}_{\beta k} = \partial_i T^{\alpha j}_{\beta k} + X_i^\gamma \Gamma^\alpha_{\gamma \nu} T^{\nu j}_{\beta k} - X_i^\gamma \Gamma^\mu_{\gamma \beta} T^{\alpha j}_{\mu k} + \Gamma^j_{im} T^{\alpha m}_{\beta k} - \Gamma^m_{ik} T^{\alpha j}_{\beta m} \tag{8}$$

where $X_i^\gamma = \partial_i X^\gamma$ is the shift tensor which reciprocally shifts space bases to surface bases, as well as space metric to surface metric; for instance, $\vec{S}_i = X_i^\alpha \vec{X}_\alpha$ and $S_{ij} = \vec{S}_i \cdot \vec{S}_j = X_i^\alpha \vec{X}_\alpha X_j^\beta \vec{X}_\beta = X_i^\alpha X_j^\beta X_{\alpha \beta}$.

Using (7, 8), one may directly prove metrilinic property of the surface metric tensor $\nabla_i S_{mn} = 0$, from where directly follows $\vec{S}_m \cdot \nabla_i \vec{S}_n = 0$, meaning that $\vec{S}_m \perp \nabla_i \vec{S}_n$ are orthogonal vectors and as so $\nabla_i \vec{S}_n$ must be parallel to the surface normal $\vec{N}$

$$\nabla_i \vec{S}_j = \vec{N} B_{ij} \tag{9}$$

$\vec{N}$ is a surface normal vector with unit length and $B_{ij}$ is the tensorial coefficient of the relationship and is generally referred as the symmetric curvature tensor. The trace of the curvature tensor with upper and lower indices is the mean curvature and its determinant is the Gaussian curvature. It is well known that a surface with constant Gaussian curvature is a sphere, consequently a sphere can be expressed as:

$$B_i^i = \lambda \tag{10}$$

where $\lambda$ is some non-zero constant. When $B_i^i = 0$ then the surface is either plane or cylinder. According to (9, 10), finding the curvature tensor defines the way of finding covariant derivatives of surface base vectors and as so, defines the way of finding surface base vectors which indirectly leads to the identification of the surface.

## 2. 2 Basics of Tensor Calculus for Moving Surfaces

All Equations provided above are generally true for moving surfaces. We now turn to a brief review of definitions of coordinate velocity $V^\alpha$, interface velocity $C$ (which is the same as normal velocity), tangent velocity $V^i$ (Figure 1), time $\dot{\nabla}$-derivatives of surface tensors and time differentiation of the surface integrals. The original definitions of time derivatives for moving surfaces were given in [27] and recently extended in [26].

Let's start from the definition of coordinate velocity $V^\alpha$ and show that the coordinate velocity is $\alpha$ component of surface velocity. Indeed,

$$V^\alpha = \frac{\partial X^\alpha}{\partial t} \tag{11}$$

On the other hand the position vector $\vec{R}$ given by (5) is tracking the coordinate particle $S^i$. Taking into account partial time differential of (5) and definition of ambient base vectors, we find

$$\vec{V} = \frac{\partial \vec{R}(t, S^i)}{\partial t} = \frac{\partial \vec{R}}{\partial X^\alpha} \frac{\partial X^\alpha(t, S^i)}{\partial t} = V^\alpha \vec{X}_\alpha \tag{12}$$

Therefore, $V^\alpha$ is ambient component of the surface velocity $\vec{V}$. Taking into account (12), normal component of the surface velocity is dot product with the surface normal, so that

$$C = \vec{V} \cdot \vec{N} = V_\alpha \vec{X}^\alpha N^\beta \vec{X}_\beta = V_\alpha N^\beta \delta^\alpha_\beta = V_\alpha N^\alpha \tag{13}$$

It is easy to show that the normal component $C$ of the coordinate velocity, generally referred as interface velocity, is invariant in contrast with coordinate velocity $V^\alpha$ and its sign depends on a choice of the normal. The projection of the surface velocity on the tangent space (Figure 1) is tangential velocity and can be expressed as

$$V^i = V^\alpha X^i_\alpha \tag{14}$$

Taking (13, 14) into account one may write surface velocity as $\vec{V} = \{C, V^i\} = C\vec{N} + V^i \vec{S}_i$. Graphical illustrations of coordinate velocity $V^\alpha$, interface velocity $C$ and tangential velocity $V^i$ are given on Figure 1. Also there is a clear geometric interpretation of the interface velocity [26]. Let the surfaces at two nearby moments of time $t$ and $t + \Delta t$ be $S_t, S_{t+\Delta t}$, correspondingly. Suppose that $A \in S_t$ (point on $S_t$) and the corresponding point $B \in S_{t+\Delta t}$, $B$ has the same surface coordinates as $A$ (Figure 2.), then $\vec{AB} \approx \vec{V} \Delta t$. Let $P$ be the point, where the unit normal $\vec{N} \in S_t$ intersects the surface $S_{t+\Delta t}$, then for small enough $\Delta t$, the angle $\angle APB \approx \pi/2$ and $AP \approx \vec{V} \cdot \vec{N} \Delta t$, therefore, $C$ can be defined as

$$C = \lim_{\Delta t \to 0} \frac{AP}{\Delta t} \tag{15}$$

and can be interpreted as the instantaneous velocity of the interface in the normal direction. It is worth of mentioning that the sign of the interface velocity depends on the choice of the normal. Although $C$ is a scalar, it is called interface velocity because the normal direction is implied.

## 2.3 Invariant Time Differentiation

Among the key definitions in calculus for moving surfaces, perhaps one of the most important one is the invariant time derivative $\dot{\nabla}$. As we have already stated above, initial preposition for time derivative was made in [27] and extended in [26]. In this paragraph, we just give geometrically intuitive definition (similar to [26]).

Suppose that invariant field $F$ is defined on the surface at all time. The idea behind the invariant time derivative is to capture the rate of change of $F$ in the normal direction. Physical explanation of why the deformations along the normal direction are so important, we give below in integration section. This is similar to how $C$ measures the rate of deformation in the normal direction. Let for a given point $A \in S_t$, find the points $B \in S_{t+\Delta t}$ and $P$ the intersection of $S_{t+\Delta t}$ and the straight line orthogonal to $S_t$ (Figure 2). Then the geometrically intuitive definition dictates that

$$\dot{\nabla} F = \lim_{\Delta t \to 0} \frac{F(P) - F(A)}{\Delta t} \tag{16}$$

As far as (16) is entirely geometric, it must be an invariant (free from choice of a reference frame). From the geometric construction one can estimate value of $F$ in point $B$, so that

$$F(B) \approx F(A) + \Delta t \frac{\partial F}{\partial t} \tag{17}$$

On the other hand, $F(B)$ is related to $F(P)$ because $B, P \in S_{t+\Delta t}$ and are nearby points on the surface $S_{t+\Delta t}$, then

$$F(B) \approx F(P) + \Delta t V^i \nabla_i F \tag{18}$$

since $\nabla_i F$ shows rate of change in $F$ along the surface and $\Delta t V^i$ captures the directed distance $BP$. Determining $F(A), F(P)$ values from (17, 18) and putting it in (16), gives

$$\dot{\nabla} F = \frac{\partial F}{\partial t} - V^i \nabla_i F \tag{19}$$

Extension of the definition (19) to any arbitrary tensors with mixed space and surface indices is given by the formula

$$\dot{\nabla} T^{\alpha i}_{\beta j} = \frac{\partial T^{\alpha i}_{\beta j}}{\partial t} - V^k \nabla_k T^{\alpha i}_{\beta j} + V^\gamma \Gamma^\alpha_{\gamma \mu} T^{\mu i}_{\beta j} - V^\gamma \Gamma^\mu_{\gamma \beta} T^{\alpha i}_{\mu j} + \dot{\Gamma}^i_k T^{\alpha k}_{\beta j} - \dot{\Gamma}^k_j T^{\alpha i}_{\beta k} \tag{20}$$

The derivative commutes with contraction, satisfies sum, product and chain rules, is metrinilic with respect to the ambient metrics and doesn't commute with the surface derivative [26]. Also from (16) it is clear that the invariant time derivative applied to time independent scalar vanishes. Christoffel symbol $\dot{\Gamma}^i_j$ for moving surfaces is defined by $\dot{\Gamma}^i_j = \nabla_j V^i - CB^i_j$.

## 2. 4 Time Differentiation of Integrals

The remarkable usefulness of the calculus of moving surfaces becomes evident from two fundamental formulas for integrations that govern the rates of change of volume and surface integrals due to the deformation of the domain [26]. For instance, in evaluation of the least action principle of the Lagrangian there is a central role for time differentiation of the surface and space integrals, from which the geometry dependence is rigorously clarified.

For any scalar field $F = F(t,S)$ defined on a Euclidean domain $\Omega$ with boundary $S$ evolving with the interface velocity $C$, the evolution of the space integral and surface integral for closed surface are given by the formulas

$$\frac{d}{dt}\int_\Omega F d\Omega = \int_\Omega \frac{\partial F}{\partial t} d\Omega + \int_S CF dS$$
$$\frac{d}{dt}\int_S F dS = \int_S \dot{\nabla} F dS - \int_S CF dS \tag{21}$$

The first term in the integral represents the rate of change of the tensor field, while the second term shows changes in the geometry. Of course there are rigorous mathematical proofs of these formulas in the tensor calculus text books. We are not going to reproduce proof of those theorems here, but instead we give less rigorous but completely intuitive explanation of why only interface velocity has to be count. Rigorous mathematical proof follows from fundamental theorem of calculus

$$\frac{d}{dt}\int_a^{b(t)} F(t,x)dx = \int_a^{b(t)} \frac{\partial F(t,x)}{\partial t} dx + b'(t)F(t,b(t)) \tag{22}$$

In the case of volume integral or surface integral it can be shown that $b'(t)$ is replaced by interface velocity $C$. Intuitive explanation is pretty simple. Propose there is no interface velocity then surface velocity only has tangent component. Tangent velocity for each given time (if there is no interface velocity) translates each point to its neighboring point and, therefore, does not add new area to the surface (or new volume to the space, or new length to the curve). As so, tangential velocity just induces rotational movement (or uniform translational motion) of the object and can be excluded from additive terms in the integration. Perhaps, it is easier to understand this statement for one dimensional motion. Let's assume that material point is moving along some trajectory (some curve), then, in each point, the velocity of the material point is tangential to the curve. Now one can translate this motion into the motion of the curve where the curve has only tangential velocity. In this aspect, the embedded curve only slides in the ambient plane (uniformly translates in the plane) without changing the length locally, therefore tangential velocity of the curve does not add new length to the curve.

## 3 RESULTS AND DISCUSSIONS

In this section we apply basics of thermodynamics and fundamental theorems of calculus of moving surfaces to demonstrate shortest proof of (4), describing motion of homogenous surface (micelle) at normal direction. We consider the system consisted of aqueous media with the formed micelle in it (Figure 3). The system is isolated with constant temperature and there is no absorbed or dissipated heat on the surface of the micelle; in other words, a process is adiabatic. We don't ask the question of how micelle forms, instead we ask why the shape of the micelle is sphere when it is thermodynamically equilibrated with the system. Strictly talking, such micelle is a subsystem of the isolated system and the surface of the micelle is closed. According to the first law of thermodynamic, as far as there is no dissipated or absorbed heat, the change of the internal energy of the surface of the micelle must be

$$dE = \delta W \tag{23}$$

where $\delta W$ is infinitesimal work done on the subsystem (micelle) and $dE$ is infinitesimal change of the internal energy. Because the temperature of the system is constant, the differential of the subsystems' internal energy can be remodeled as

$$dE = dU \tag{24}$$

where $U$ is the total potential energy of the micelle. By the definition the elementary work done on the subsystem is

$$\delta W = (P^- + \Pi) d\Omega \tag{25}$$

where $P^-, \Pi$ are external hydrodynamic and osmotic pressures applied on the surface of the micelle by the surroundings correspondingly and $\Omega$ is the volume of the micelle with boundary of $S$ surface area. Now let's propose that the surface of the micelle is homogenous and can be described by Laplace model of surface tension, then

$$dU = \sigma dS \tag{26}$$

$\sigma$ is surface tension. As far as we discuss simplest case of the system consisted of aqueous medium and single micelle, we can suggest that the surface tension is not time variable. Using (23-26) after few lines of algebra

$$\int_S \sigma dS = \int_\Omega (P^- + \Pi) d\Omega \tag{27}$$

Assuming the surface of the micelle is moving so that (27) stays valid for any time variations, then time differentiation of the left side must be equal to time differentiation of right integral. As far as on the right hand side we have space integral, time differentiation can be taken into the integral (using (21)), so that integration theorem for space integral holds and the convective and advective terms due to volume motion are considered

$$\frac{d}{dt}\int_\Omega (P^- + \Pi) d\Omega = \int_\Omega (\partial_\alpha P^- + \partial_\alpha \Pi)\frac{\partial X^\alpha}{\partial t} d\Omega + \int_S C(P^- + \Pi) dS \tag{28}$$

To calculate time derivative of the surface integral we have to take into account the theorem about time differentiation of the surface integral (21), from which follows that for time invariable surface tension

$$\frac{d}{dt}\int_S \sigma dS = \int_S -\sigma C B_i^i dS \tag{29}$$

Where $C = V^\alpha N_\alpha$ is interface velocity, $N_\alpha$ is $\alpha$ component of the surface normal and $V^\alpha = \partial X^\alpha / \partial t$ is coordinate velocity, $X^\alpha$ is general coordinate and $B_i^i$ is the trace of the mixed curvature tensor generally known as mean curvature. After few lines of algebra putting (27-29) together

$$\int_S \left(\sigma C B_i^i + C(P^- + \Pi)\right) dS = -\int_\Omega (\partial_\alpha P^- + \partial_\alpha \Pi) V^\alpha d\Omega \qquad (30)$$

Generalized Gauss theorem converts the surface integral of the left hand side of (30) into space integral, so that

$$\int_S N_\alpha V^\alpha \left(\sigma B_i^i + (P^- + \Pi)\right) dS = \int_\Omega \partial_\alpha (\sigma V^\alpha B_i^i + (P^- + \Pi) V^\alpha) d\Omega \qquad (31)$$

Combination of (30) and (31) immediately gives equation of motion for the micelle surface

$$\partial_\alpha (\sigma V^\alpha B_i^i + (P^- + \Pi) V^\alpha) = -(\partial_\alpha P^- + \partial_\alpha \Pi) V^\alpha \qquad (32)$$

For equilibrium processes internal and external pressures are identical $P^- = P^+$, so that (32) becomes identical to (4). Also, we should note that (32) is only valid for motion of the homogeneous surfaces at normal direction, therefore, it doesn't display any deformation in tangent directions. (32) further simplifies when the micelle comes in equilibrium with the solvent where divergence of the surface velocity $\partial_\alpha V^\alpha$ (stationary shape) along with $\partial P / \partial t$ (where $P = P^- + \Pi$) vanishes then the solution to (32), taking into account the condition (28), becomes

$$B_i^i = -\frac{P}{\sigma} \qquad (33)$$

Incidentally, stationary solution $\partial_\alpha V^\alpha = 0$ to (32) dictates $B_i^i = 0$ corresponding to cylindrical and lamellar surfaces. The result (33) shows that the solution is surfaces which have constant mean curvatures (CMC). Such surfaces are rare and can be many if one relaxes the condition we restricted the system. Namely we consider isolated system where micelle is closed sub-system, these two preconditions mathematically mean that the micelle surface we discuss is compact embedded surface in $\mathbb{R}^3$. According to A. D. Alexandrov uniqueness theorem for surfaces, a compact embedded surface in $\mathbb{R}^3$ with constant non-zero mean curvature is a sphere (A. D. Alexandrov Amer. Math. Soc. Trans. 21, 412 (1958)). Correspondingly the solution (33) is a sphere (as far as we have compact two-manifold in the Euclidian space). Therefore, when

$$\frac{P}{\sigma} \neq 0 \qquad (34)$$

the micelle is spheroid and becomes lamella or cylinder when the pressure along the surface over the surface tension vanishes. This surprisingly simple and elegant derivation explains all the shapes of micelles in aqueous solution in equilibrium conditions. Furthermore if the compactness condition is relaxed then (33) predicts that all other CMC surfaces are also possible. If one takes into account that the surface tension in general can be a function of many variables, such as Gaussian curvature, bending rigidity, spontaneous curvature, lipid concentration and etc., than (32) may predict possible deformations of differently shaped micelles and their wide range of static shapes. In fact, if considered that surface tension, which is defined as potential energy per unit area, can be a function of mean curvature $\sigma = \sigma(B_i^i)$, then Taylor expansion of $\sigma(B_i^i)$ naturally rises terms related to Gaussian curvature, spontaneous curvature, bending rigidity etc. Of course all these generalizations along with taking into account temperature fluctuations can be included in the equations, which we won't be doing in this paper, because unfavorable complication of already complicated equations should be avoided and it should be a source for another paper.

Based on (33) we can calculate minimal value of a micelle radius. The value of the trace of the mixed curvature tensor for a sphere is

$$B_i^i = -\frac{2}{R} \qquad (35)$$

$R$ is radius of the micelle. Let's calculate value of the surface pressure when the micelle still can exist. Lipids in a micelle are confined in the surface by hydrophobic interactions with average energy in the range of hydrogen bonding. As far as values of hydrogen bonding energy are somewhat uncertain in the literature, by the first approximation we take average energy for the hydrogen bonding energy interval and assign it to the lipid molecule. Low boundary of the interval (minimum energy) for $XH \cdots Y$ hydrogen bond is about 1 kJ per mol ($CH \cdots C$ unit) and high boundary is about 161 kJ per mol ($FH \cdots F$ unit), the low and high values are taken according to references [28, 29]. Therefore, average energy is about $(1+161)/2 = 81 kJ/mol \approx 13 \cdot 10^{-20} J$. To estimate hydrogen bonding energy per molecule with the undefined shape (lipid molecule) in the first approximation is to assign average energy to it and consider the spherical shape with the gyration radius. Of course it is low level approximation, but even such rough calculations produces reasonable results. After all these rough estimations the pressure to move one lipid from the surface, in order to induce critical deformations of the surface, is about average energy per the average volume of the lipid molecule

$$P \approx 3 \cdot 13 \cdot 10^{-20} / 4\pi r_G^3 \approx 3.1 \cdot 10^7 \, N/m^2 \qquad (36)$$

where $4\pi r_G^3/3$ is the estimated volume of a lipid molecule considered as sphere with the gyration radius $r_G \approx 1 nm$. On the other hand, surface tension of a fluid monolayer at optimal packing of the lipids is about $\sigma \approx 3 \cdot 10^{-2} N/m$ [30-32], using these and (35, 36) in (33) the estimated micelle radius is

$$R \approx \frac{2 \cdot 3 \cdot 10^{-2}}{3.1 \cdot 10^7} = (19.3 \pm 0.1) \, \overset{0}{\text{A}} \qquad (37)$$

These calculations put the minimum radius of the micelle in nanometer scale and is in very good agreement with experimental as well as computational frameworks [33, 34]. To further validate the (37) result, we ran a CHARMM based Micelle Builder simulation [35, 36] for 100 phospholipid molecules (1,2-Dimyristoyl-sn-Glycero-3-Phosphocholine, DHPC). The simulation result (Figure 4) generated a spherical micelle with diameter $(38.5 \pm 0.1)$ Å. These calculations indeed indicate that even such rough estimations produce reasonable accuracy.

Of course, the first approximation is low level. To get more convincing estimations it is necessary to take into account that neither lipids are spherical nor hydrophobic interactions per lipid are average energy of single hydrogen bond. The simulation results discussed above (so as calculations) were done on the first approximation, where lipids were considered as spherical and hydrophobic energy per lipid was estimated as the single hydrogen bonding energy. To produce higher level approximation and the comparison with the theory, we demonstrate all atom simulation and estimation of the radius in the second approximation, where lipids are no longer undefined spheres, but have well defined surfactant geometry and hydrophobic energy is no longer average energy of single hydrogen bond, but is 1 kJ per mol per $-CH_2-$ unit. In simulations we used DHPC lipid molecule having 12 $-CH_2-$ units (Figure 5) per hydrophobic tail, so hydrophobic energy is about $12 kJ/mol \approx 1.99 \cdot 10^{-20} J$. Accurate calculation of the lipid molecule volume using cavity, channel and cleft volume calculator [37], gives the volume estimation of about 894 Å$^3$. Using this value, one gets

$$P \approx \frac{1.99 \cdot 10^{-20}}{0.894 \cdot 10^{-27}} \approx 2.22 \cdot 10^7 \, N/m^2 \tag{38}$$

On the other hand, using the same surface tension of a fluid monolayer at the optimal packing of the lipids, one gets

$$R \approx \frac{2 \cdot 3 \cdot 10^{-2}}{2.22 \cdot 10^7} = (27 \pm 0.1) \, \overset{0}{\text{Å}} \tag{39}$$

All atom simulation also generates spherical structure with diameter $(54 \pm 0.1) \overset{0}{\text{Å}}$ (Figure 5), although there is still some uncertainty in this estimation because we assigned 1 kJ/mol energy per $-CH_2-$ unit and we based on references [28, 29] data, while, for instance, in [38] it is mentioned that the hydrophobic interactions are about 4 kJ/mol per $-CH_2-$ unit. In our opinion, this discrepancy can be resolved if one calculates hydrophobic energy based on the potential energy

$$U_{CH_2} = -\int_\Omega \frac{\varepsilon_0}{2} (E_{CH_2})^2 d\Omega \tag{40}$$

where $\vec{E}_{CH_2}$ is electric field per $-CH_2-$, $\varepsilon_0$ is dielectric constant in the vacuum and $\Omega$ stands for the volume of the lipid molecules. (40) directly emerges from $F_{\mu\nu}F^{\mu\nu}$ term written in (1). In fact, for electrostatics

$$U = \int_\Omega \frac{1}{\mu_0} F_{\mu\nu} F^{\mu\nu} d\Omega = -\int_\Omega \frac{\varepsilon_0}{2} E^2 d\Omega \tag{41}$$

and one should go to the scrutiny of calculating electric field for each $-CH_2-$ units, then take a sum of the electric field and square it (we are not going to do it in this paper). Also, one may ask why the hydrophilic interaction energy is not taken into account in these calculations. Hydrophilic head of the lipid molecule is in contact with water molecules and does not need any work to be done to drag it in aqueous solution from the lipids layer. Therefore, hydrophilic interaction energy can be neglected. The most of the work goes on overcoming hydrophobic interactions between lipid tails.

## 4 CONCLUSIONS

To summarize, we have presented a framework for the analysis of micelle dynamics using first law of thermodynamics and calculus of moving surfaces. Final equations of normal motion (4, 32) are based on the assumption that the micelle surface is homogeneous and is restricted by precondition to the surfaces, which can be described by time invariable surface tension. However, (1, 2) don't have homogeneity constrain and indicate motion along normal deformation, as well as deformation into tangent directions, but are analytically more complex. The solution to the normal equations of motion in equilibrium conditions turned to be surprisingly simple and displayed all possible equilibrium shapes of micelles. Micelles are spheroids and become lamellas or cylinders when the pressure along the surface over the surface tension vanishes. The proposed formalism was illustrated by applying it to the estimation of micelle optimal radius and comparison to all atom simulations. The remarkable accuracy was found even for low level approximations between theoretically calculated radius and the one obtained from the atomistic simulations. As a final remark, the proposed theory can be readily extended to any homogenous surfaces, such are vesicles and membranes.

## ACKNOWLEDGMENTS

We were partially supported by the personal savings accumulated during visits to the Department of Mechanical Engineering, Department of Chemical Engineering, OCMB Philips Institute and Institute for Structural Biology and Drug Discovery of Virginia Commonwealth University in 2007-2012 years. Limited access to Virginia Commonwealth University's library in 2012-2013 years is also gratefully acknowledged.

## APPENDIX

Here, we derive exact equations (2) for moving surfaces, where in "exact" we mean that there have not been done any approximations while evaluating kinetic and potential energies of the surface. Beforehand, we should mention that (2) also follows from (1) if one applies same formalism as it is given in (23-25). Indeed, in non-relativistic framework space is three-dimensional Euclidian, the surface is two-dimensional Riemannian and potential energy becomes

$$U = \int_\Omega (\frac{1}{\mu_0} F_{\mu\nu} F^{\mu\nu} + A_\alpha J^\alpha) d\Omega = \int_\Omega \frac{1}{2}(-\varepsilon_0 E^2 + \frac{1}{\mu_0} B^2) + q\varphi - \vec{A}\vec{J}) d\Omega \qquad (42)$$

Where $\vec{E}, \vec{B}$ are electric and magnetic fields and $q, \varphi, \vec{A}, \vec{J}$ are charge density, electric potential, magnetic vector potential and current density vector respectively. Using (23-25) one gets $dU = -(P^+ + \Pi)d\Omega$, $-(P^+ + \Pi) = 1/\mu_0 F_{\mu\nu} F^{\mu\nu} + A_\alpha J^\alpha$ and taking into account that the pressure comes from the "normal force" applied to the surface $-\partial_\alpha (P^+ + \Pi)V^\alpha = fa$, and in tangent direction $f_i a^i = 0$, then (1) becomes (2). It might be more helpful providing details about (1), but, as far as it goes out of the scope of this paper, we are reluctant to do it here.

Now we turn to the derivation of (2) without using any information from (1). To deduce the equations of motion we deduce the simplest one from the set (2) first. It is direct consequence of generalization of conservation of mass low. Variation of the surface mass density must be so that $dm/dt = 0$, where $m = \int_S \rho dS$ is surface mass with $\rho$ surface density. Since the surface is closed, at the boundary condition $v = n_i V^i = 0$, a pass integral along any curve $\gamma$ across the surface must vanish ($n_i$ is a normal of the curve and lays in the tangent space). Using Gauss theorem, conservation of mass and integration formula (21), we find

$$0 = \int_\gamma v\rho d\gamma = \int_\gamma n_i V^i \rho d\gamma = \int_S \nabla_i(\rho V^i)dS =$$
$$= \int_S (\nabla_i(\rho V^i) - \rho CB_i^i + \rho CB_i^i)dS =$$
$$= \int_S (\nabla_i(\rho V^i) - \rho CB_i^i)dS + \int_S \dot{\nabla}\rho dS - \frac{d}{dt}\int_S \rho dS = \quad (43)$$
$$= \int_S (\dot{\nabla}\rho + \nabla_i(\rho V^i) - \rho CB_i^i)dS$$

Since last integral mast be identical to zero for any integrand, one immediately finds first equation from the set (2). To deduce second and third equations, we take a Lagrangian

$$L = \int_S \frac{\rho V^2}{2}dS + \int_\Omega (P^+ + \Pi)d\Omega \quad (44)$$

and set minimum action principle requesting that $\delta L/\delta t = 0$. Evaluation of space integral is simple and straightforward, using integration theorem for space integral where the convective and advective terms due to volume motion is properly taken into account (21), we find

$$\frac{\delta}{\delta t}\int_\Omega (P^+ + \Pi)d\Omega = \int_\Omega \partial_\alpha (P^+ + \Pi)V^\alpha d\Omega + \int_S C(P^+ + \Pi)dS \quad (45)$$

Derivation for kinetic part is a bit tricky and challenging that is why we do it last. Straightforward, brute mathematical manipulations, using first equation from (2), lead

$$\frac{\delta}{\delta t}\int_S \frac{\rho V^2}{2}dS = \int_S (\dot{\nabla}\frac{\rho V^2}{2} - CB_i^i \frac{\rho V^2}{2})dS = \int_S (\dot{\nabla}\rho \frac{V^2}{2} + \rho \dot{\nabla}\frac{V^2}{2} - CB_i^i \frac{\rho V^2}{2})dS =$$
$$= \int_S (\rho CB_i^i - \nabla_i(\rho V^i))\frac{V^2}{2} + \rho \dot{\nabla}\frac{V^2}{2} - CB_i^i \frac{\rho V^2}{2})dS = \int_S (-\nabla_i(\rho V^i)\frac{V^2}{2} + \rho \dot{\nabla}\frac{V^2}{2})dS =$$
$$= \int_S (-\nabla_i(\rho V^i \frac{V^2}{2}) + \rho V^i \nabla_i \frac{V^2}{2} + \rho \dot{\nabla}\frac{V^2}{2})dS = \quad (46)$$
$$= \int_S (-\nabla_i(\rho V^i \frac{V^2}{2}) + \rho V^i \nabla_i \frac{\vec{V}\cdot\vec{V}}{2} + \rho \dot{\nabla}\frac{\vec{V}\cdot\vec{V}}{2})dS =$$
$$= \int_S [-\nabla_i(\rho V^i \frac{V^2}{2}) + \rho \vec{V}(V^i \nabla_i \vec{V} + \dot{\nabla}\vec{V})]dS$$

At the end point of variations the surface reaches stationary point and therefore by Gauss theorem (as we used it already in (43)), we find

$$\int_S -\nabla_i(\rho V^i \frac{V^2}{2}) dS = -\int_\gamma \rho V^i n_i \frac{V^2}{2} d\gamma = 0 \qquad (47)$$

$\gamma$ is stationary contour of the surface and $n_i$ is the normal to the contour, therefore interface velocity for contour $v = n_i V^i = 0$ and the integral (47) vanishes, correspondingly

$$\frac{\delta}{\delta t} \int_S \frac{\rho V^2}{2} dS = \int_S \rho \vec{V}(V^i \nabla_i \vec{V} + \dot{\nabla}\vec{V}) dS \qquad (48)$$

To decompose dot product in the integral by normal and tangential components and, therefore, deduce final equations, we do following algebraic manipulations

$$\dot{\nabla}\vec{V} + V^i \nabla_i \vec{V} = \dot{\nabla}\vec{V} + V^i \nabla_i \vec{V} + CV^i B_i^j \vec{S}_j - CV^i B_i^j \vec{S}_j = \\ = \dot{\nabla}\vec{V} + V^i \nabla_i \vec{V} + CV^i B_i^j X_j^\alpha \vec{X}_\alpha - CV^i B_i^j \vec{S}_j \qquad (49)$$

Now using Weingarten's formula $X_j^\alpha B_i^j = -\nabla_i N^\alpha$, metrinilic property of Euclidian space base vectors $\nabla_i \vec{X}_\alpha = 0$ and definition of surface normal $\vec{N} = N^\alpha \vec{X}_\alpha$ last equation transforms

$$\dot{\nabla}\vec{V} + V^i \nabla_i \vec{V} + CV^i B_i^j X_j^\alpha \vec{X}_\alpha - CV^i B_i^j \vec{S}_j = \\ = \dot{\nabla}\vec{V} + V^i \nabla_i \vec{V} - CV^i \vec{X}_\alpha \nabla_i N^\alpha - CV^i B_i^j \vec{S}_j = \\ = \dot{\nabla}\vec{V} + V^i \nabla_i \vec{V} - CV^i \nabla_i (N^\alpha \vec{X}_\alpha) - CV^i B_i^j \vec{S}_j = \\ = \dot{\nabla}\vec{V} + V^i \nabla_i \vec{V} - CV^i \nabla_i \vec{N} - CV^i B_i^j \vec{S}_j \qquad (50)$$

Taking into account $\vec{V} = C\vec{N} + V^i \vec{S}_i$, $\nabla_i \vec{V} = \nabla_i (C\vec{N}) + \nabla_i (V^j \vec{S}_j)$ and $\dot{\nabla}\vec{V} = \dot{\nabla}(C\vec{N}) + \dot{\nabla}(V^j \vec{S}_j)$ we have

$$\dot{\nabla}\vec{V} + V^i \nabla_i \vec{V} - CV^i \nabla_i \vec{N} - CV^i B_i^j \vec{S}_j = \\ = \dot{\nabla}\vec{V} + V^i \nabla_i (C\vec{N}) + V^i \nabla_i (V^j \vec{S}_j) - CV^i \nabla_i \vec{N} - CV^i B_i^j \vec{S}_j = \\ = \dot{\nabla}\vec{V} + V^i \vec{N} \nabla_i C + V^i \nabla_i (V^j \vec{S}_j) - CV^i B_i^j \vec{S}_j = \\ = \dot{\nabla}(C\vec{N}) + \dot{\nabla}(V^j \vec{S}_j) + V^i \vec{N} \nabla_i C + V^i \nabla_i (V^j \vec{S}_j) - CV^i B_i^j \vec{S}_j \qquad (51)$$

Continuing algebraic manipulations using Thomas formula $\dot{\nabla}\vec{N} = -\nabla^i C \vec{S}_i$, the formula for surface derivative of interface velocity $\vec{N}\nabla_i C = \dot{\nabla}\vec{S}_i$ and the definition of curvature tensor $\vec{N} B_{ij} = \nabla_i \vec{S}_j$ yield

$$
\begin{aligned}
&\dot{\vec{\nabla}}(C\vec{N}) + \dot{\vec{\nabla}}(V^j\vec{S}_j) + V^i\vec{N}\nabla_i C + V^i\nabla_i(V^j\vec{S}_j) - CV^i B_i^j\vec{S}_j = \\
&= \dot{\vec{\nabla}}(C\vec{N}) + C\nabla^j C\vec{S}_j + 2V^i\vec{N}\nabla_i C + V^iV^j B_{ij}\vec{N} + \dot{\vec{\nabla}}(V^j\vec{S}_j) - \\
&\quad - V^i\vec{N}\nabla_i C + V^i\nabla_i(V^j\vec{S}_j) - V^iV^j B_{ij}\vec{N} - C\nabla^j C\vec{S}_j - CV^i B_i^j\vec{S}_j = \\
&= \dot{\vec{\nabla}}(C\vec{N}) - C\dot{\vec{\nabla}}\vec{N} + 2V^i\vec{N}\nabla_i C + V^iV^j B_{ij}\vec{N} + \dot{\vec{\nabla}}(V^j\vec{S}_j) - \\
&\quad - V^j\dot{\vec{\nabla}}\vec{S}_j + V^i\nabla_i(V^j\vec{S}_j) - V^iV^j\nabla_i\vec{S}_j - C\nabla^j C\vec{S}_j - CV^i B_i^j\vec{S}_j = \\
&= (\dot{\vec{\nabla}} C + 2V^i\nabla_i C + V^iV^j B_{ij})\vec{N} + (\dot{\vec{\nabla}} V^j + V^i\nabla_i V^j - C\nabla^j C - CV^i B_i^j)\vec{S}_j
\end{aligned}
\qquad (52)
$$

Doting (52) on $\vec{V}$ and combining it with (48) last derivation finally reveals variation of kinetic energy, so that we finally get

$$
\begin{aligned}
\frac{\delta}{\delta t}\int_S \frac{\rho V^2}{2} dS &= \int_S \rho\vec{V}(V^i\nabla_i\vec{V} + \dot{\vec{\nabla}}\vec{V}) dS = \\
&= \int_S \begin{pmatrix} \rho C(\dot{\vec{\nabla}} C + 2V^i\nabla_i C + V^iV^j B_{ij}) + \\ \rho V_i(\dot{\vec{\nabla}} V^i + V^j\nabla_j V^i - C\nabla^i C - CV^j B_j^i) \end{pmatrix} dS
\end{aligned}
\qquad (53)
$$

Combining (43-45) and (53) together and taking into account that the pressure acts on the surface along the surface normal, we immediately find first and the last equation of the (2). To clarify second equation, we have

$$
\begin{aligned}
\int_S \rho C\left(\dot{\vec{\nabla}} C + 2V^i\nabla_i C + B_{ij}V^iV^j\right) dS &= \int_\Omega -\partial_\alpha(P^+ + \Pi)\cdot V^\alpha d\Omega - \int_S C(P^+ + \Pi) dS \\
\int_S \rho C\left(\dot{\vec{\nabla}} C + 2V^i\nabla_i C + B_{ij}V^iV^j + (P^+ + \Pi)\right) dS &= \int_\Omega -\partial_\alpha(P^+ + \Pi)\cdot V^\alpha d\Omega
\end{aligned}
\qquad (54)
$$

After applying Gauss theorem to the second equation the surface integral is converted to space integral so that one gets

$$
\partial_\alpha\left(\rho V^\alpha\left(\dot{\vec{\nabla}} C + 2V^i\nabla_i C + B_{ij}V^iV^j\right) + (P^+ + \Pi)V^\alpha\right) = -\partial_\alpha(P^+ + \Pi)\cdot V^\alpha
\qquad (55)
$$

and, therefore, all three equations of (2) are rigorously clarified.

**REFERENCES**

[1] C. Tanford, *The Hydrophobic Effect Formation of Micelles and Biological Membranes* (Wiley-Interscience, New York, 1973).

[2] M. Deserno, Chemistry and Physics of Lipids **185**, 11 (2015).

[3] U. Seifert, and R. Lipowsky, Chapter 8: Morphology of Vesicles. In: Lipowsky, R., Sackmann, E. (Eds.), Structure and Dynamics of Membranes; vol. 1 of Handbook of Biological Physics (North-Holland, Amsterdam, 1995) pp. 403-463.

[4] U. Seifert, Adv. Phys. 46, 13 (1997).

[5] P. B. Canham, J. Theor. Biol. **26,** 61 (1970).

[6] W. Helfrich, Z. Naturforsch. C **28,** 693 (1973).

[7] E. A. Evans, Biophys. J. **14,** 923 (1974).

[8] Z.C. Ou-Yang, and W. Helfrich, Phys. Rev. Lett. **59,** 2486 (1987).

[9] Z.C. Ou-Yang, and W. Helfrich, Phys. Rev. A **39,** 5280 (1989).

[10] S. Svetina, and B. Zeks, Eur. Biophys. J. **17,** 101 (1989).

[11] U. Seifert, and R. Lipowsky, Phys. Rev. A **42,** 4768 (1990).

[12] R. Lipowsky, Nature **349,** 475 (1991).

[13] U. Seifert, K. Berndl, and R. Lipowsky, Phys. Rev. A **44,** 1182 (1991).

[14] F. Jülicher, and R. Lipowsky, Phys. Rev. Lett. **70,** 2964 (1993).

[15] F. Jülicher, and R. Lipowsky, Phys. Rev. E **53,** 2670 (1996).

[16] F. Jülicher, and U. Seifert, Phys. Rev. E **49,** 4728 (1994).

[17] L. Miao, U. Seifert, M. Wortis, and H. G. Döbereiner, Phys. Rev. E **49,** 5389 (1994).

[18] V. Heinrich, S. Svetina, and B. Zeks, Phys. Rev. E **48,** 3112 (1993).

[19] V. Kralj-Iglic, S. Svetina, and B. Zeks. Eur. Biophys. J. **22,** 97 (1993).

[20] D.V. Svintradze, Biophys. J. **108,** 512 (2015).

[21] D. Chandler, Nature **437,** 640 (2005).

[22] S. Leikin, V. A. Parsegian, and D. C. Rau, Ann. Rev. Phys. Chem. **44,** 369 (1993).

[23] D.V. Svintradze, Biophys. J. **98,** 43 (2010)

[24] D.V. Svintradze, Biophys. J. **110,** 623 (2016).

[25] P. Grinfeld, J. Geom. Symm. Phys. **16,** 1 (2009).

[26] P. Grinfeld, *Introduction to Tensor Analyses and the Calculus of Moving Surfaces* (Springer, New York, 2010).

[27] J. Hadamard, *Mmoire sur le problme d'analyse relatif l'quilibre des plaques elastiques encastres* (Oeuvres, Tome 2. Hermann, 1968).

[28] J. W. Larson, and T. B. McMahon, Inor. Chem. **23,** 2029 (1984).

[29] J. Emsley, Chem. Soc. Rev. **9,** 91 (1980).

[30] F. Jahnig, Biophys. J. **71,** 1348 (1996).

[31] J. N. Israelachvili, D. J. Mitchell, and B. W. Ninham, Biochim. Biophys. Acta. **470,** 185 (1977).

[32] J. N. Israelachvili, *Intermolecular and Surface Forces: Revised Third Edition* (Academic Press, London, 2011).

[33] S. E. Feller, Y. Zhang, and R. W. Pastor, J. Chem. Phys. **103,** 10267 (1995).

[34] E. Egberts, S. J. Marrink, and H. J. C. Berendsen, Eur. Biophys. J. **22,** 423 (1994).

[35] S. Jo, T. Kim, V.G. Iyer, and W. Im. J. Comput. Chem. **29,** 1859 (2008).

[36] X. Cheng, S. Jo, H. S. Lee, J. B. Klauda and W. Im, J. Chem. Inf. Model. **53,** 2171 (2013).

[37] N. R. Voss, and M. Gerstein, Nucleic Acids Res. **38,** W555 (2010).

[38] A. Leitmannova Liu, *Volume 4. Advances in Planar Lipids Bilayers and Liposomes* (Academic Press. Elsevier, 2011).

**FIGURE LEGENDS**

FIGURE 1. Graphical illustration of the arbitrary surface and its' local tangent plane. $\vec{S}_1, \vec{S}_2, \vec{N}$ are local tangent plane base vectors and local surface normal respectively. $\vec{X}_1, \vec{X}_2, \vec{X}_3$ are arbitrary base vectors of the ambient Euclidean space and $\vec{R} = \vec{R}(X) = \vec{R}(t,S)$ is radius vector of the point. $\vec{V}$ is arbitrary surface velocity and $C, V_1, V_2$ display projection of the velocity to the $\vec{N}, \vec{S}_1, \vec{S}_2$ directions respectively.

FIGURE 2. Geometric interpretation of the interface velocity $C$ and of the curvilinear time derivative $\dot{\nabla}$ applied to invariant field $F$. $A$ is arbitrary chosen point so that it lays on $F(S_t) \in S_t$ curve and $B$ is its' corresponding point on the $S_{t+\Delta t}$ surface. $P$ is the point where $S_t$ surface normal, applied on the point $A$, intersects the surface $S_{t+\Delta t}$. By the geometric construction, for small enough $\Delta t \to 0$, $\angle APB \to \pi/2$, $\vec{AB} \approx \vec{V}\Delta t$ and $AP \approx \vec{V}\cdot\vec{N}\Delta t$. On other hand, by the same geometric construction the field $F$ in the point $B$ can be estimated as $F(B) \approx F(A) + \Delta t \partial F / \partial t$, while from viewpoint of the $S_{t+\Delta t}$ surface the $F(B)$ value can be estimated as $F(P) + \Delta t V^i \nabla_i F$, where $\nabla_i F$ shows rate of change in $F$ along the surface $S_{t+\Delta t}$ and along the directed distance $BP \approx \Delta t V^i$.

FIGURE 3. (Color online) Graphical illustration of the isolated system containing aqueous solution. Water molecules are represented as red and white sticks. The system boundary is shown as white faces with black edges. The subsystem-micelle is closed surface, blue blob in the center of the system.

FIGURE 4. (Color online) Simulated three dimensional coordinates of the micelle in aqueous solution display sphere with diameter 38,5Å. (Left) 1,2-Dimyristoyl-sn-Glycero-3-Phosphocholine molecules (DHPC phospholipids) are modeled as orange balls. (Right) Gaussian mapping at contour resolution 8Å of the micelle shows spherical structure.

FIGURE 5. (Color online) All atom simulation of a micelle consisted from 1,2-Dimyristoyl-sn-Glycero-3-Phosphocholine molecules. (A) The figure shows a geometry of the DHPC surfactant molecule used in simulation and gives parametric description of volume, surface area, sphericity and effective radius. (B) Indicates atomistic simulation result contoured by Gaussian map and the diameter of the micelle, measured by PyMol. The diameter of the simulated micelle appears to be 54,0 Å with the uncertainty of the measurement 0,1 Å.

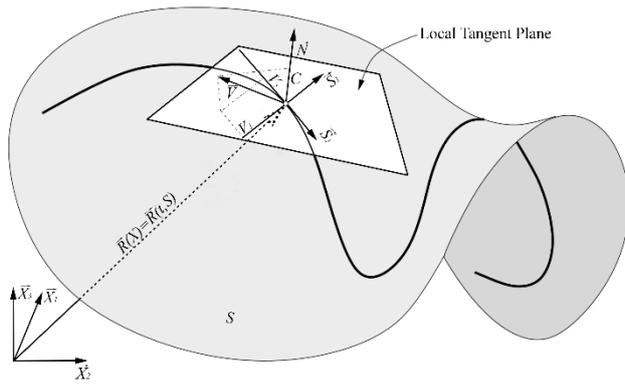

Figure 1.

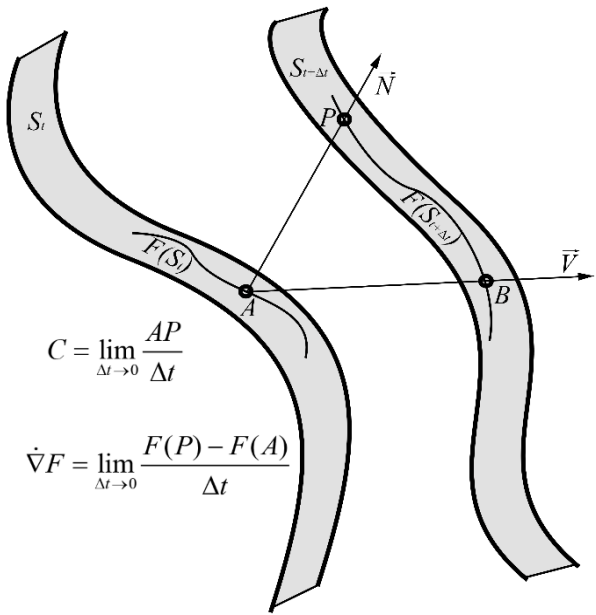

Figure 2.

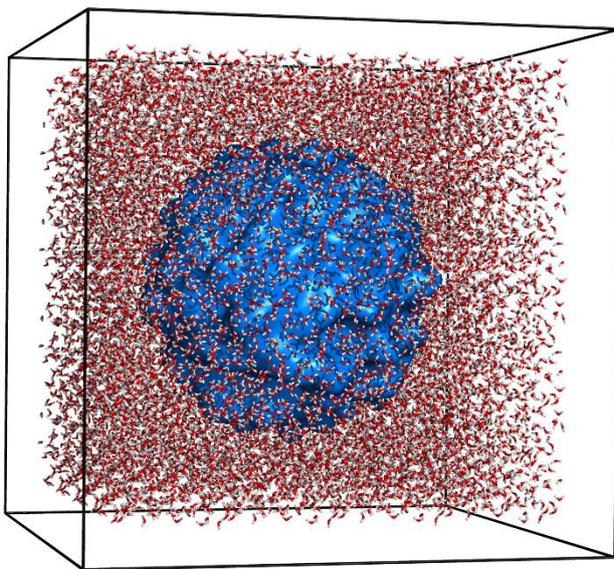

Figure 3.

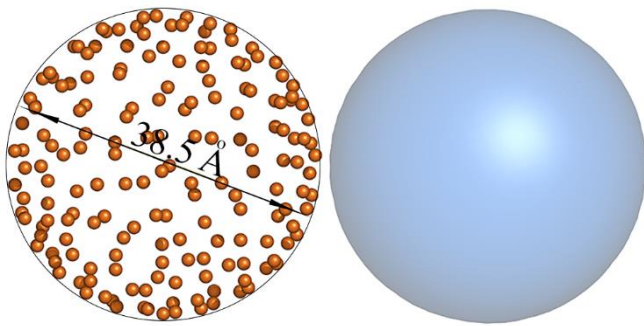

Figure 4.

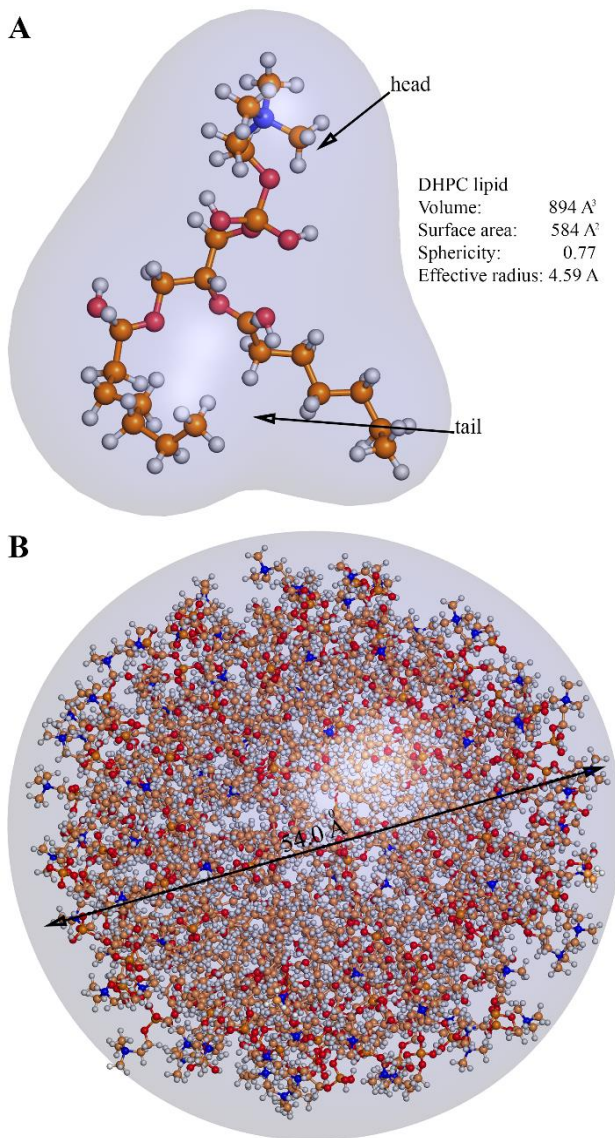

Figure 5.